\documentclass[10pt,conference,letterpaper]{IEEEtran}
\IEEEoverridecommandlockouts
\pdfoutput=1
\usepackage{cite}
\usepackage{amsmath,amssymb,amsfonts}
\usepackage{algorithm}
\usepackage[noend]{algpseudocode}
\usepackage{caption}
\usepackage{subcaption}
\usepackage{graphicx}
\usepackage{textcomp}
\usepackage{dcolumn}% Align table columns on decimal point
\usepackage{bm}% bold math
\usepackage{braket}
\usepackage{svg}
\usepackage{hyperref}% add hypertext capabilities 
\hypersetup{
    colorlinks = true,
    citecolor = magenta,
    linkcolor = purple
}
\def\BibTeX{{\rm B\kern-.05em{\sc i\kern-.025em b}\kern-.08em
    T\kern-.1667em\lower.7ex\hbox{E}\kern-.125emX}}

\usepackage{adjustbox}
\usepackage{flushend}
\usepackage{siunitx}
\usepackage{booktabs}
\usepackage{tabularx}
\usepackage{qcircuit}
\usepackage{svg}
\usepackage{xspace}
\newtheorem{innercustomgeneric}{\customgenericname}
\providecommand{\customgenericname}{}
\newcommand{\newcustomtheorem}[2]{%
  \newenvironment{#1}[1]
  {%
   \renewcommand\customgenericname{#2}%
   \renewcommand\theinnercustomgeneric{##1}%
   \innercustomgeneric
  }
  {\endinnercustomgeneric}
}

\newcustomtheorem{customthm}{Theorem}
\newcustomtheorem{definitionBob}{Definition}

\newcommand*\chem[1]{\ensuremath{\mathrm{#1}}}

\newcommand{\scross}{\textsc{Cross}\xspace}
\newcommand{\sbar}{\textsc{Bar}\xspace}

\newcommand{\sw}[2]{\textsc{#1}$_{#2}$}

\usepackage{listings}

\definecolor{commentstyle}{HTML}{88846f}
\definecolor{numberstyle}{HTML}{ae81ff}
\definecolor{stringstyle}{HTML}{e6db64}
\definecolor{backcolour}{HTML}{272822}
\definecolor{keywordstyle}{HTML}{f92672}
\lstdefinestyle{mystyle}{
    backgroundcolor=\color{backcolour},  % Background color
    commentstyle=\color{commentstyle}, % comment color
    keywordstyle=\color{keywordstyle}, % e.g. import from for
    stringstyle=\color{stringstyle}, % color of string in python
    identifierstyle=\color{white}, % variable color
    numberstyle=\color{commentstyle}, % <- not responded
    basicstyle=\ttfamily\footnotesize\color{numberstyle}, % bracket, number related
    rulecolor=\color{blue},  
    breakatwhitespace=false,         
    breaklines=true,                 
    captionpos=b,                    
    keepspaces=false,                 
    numbers=left,                    
    numbersep=5pt,                  
    showspaces=false,                
    showstringspaces=false,
    showtabs=false,                  
    tabsize=2
}

\usepackage{xcolor}
\usepackage{tabularx, makecell, linegoal}

\usepackage{ifthen}

\newboolean{ShowComments}
\setboolean{ShowComments}{true}  % change to true to show remaining colorful author comments. 
\ifthenelse{\boolean{ShowComments}}%
	{
		\newcommand{\ColorComment}[3]{%
				{\colorbox{#1}{\color{white}   \textsf{\textbf{#2}}} \textcolor{#1}{#3}}}%  Colorful box, initials, phrase 

	}%
	{
		\newcommand{\ColorComment}[3]{}%  Do nothing at all

	}%

\definecolor{rdvcolor}{rgb}{0,0.5,0}\newcommand{\rdv}[1]{\ColorComment{rdvcolor}{rdv}{#1}}
\definecolor{satohcolor}{RGB}{254,0,0}
\definecolor{michalcolor}{RGB}{255,127,80}
\definecolor{naphanncolor}{RGB}{112, 51, 173}
\definecolor{sitongcolor}{rgb}{0,0.5,0}
\definecolor{simoncolor}{rgb}{0,0.5,0}
\definecolor{amincolor}{RGB}{96, 189, 239}
\definecolor{shotacolor}{RGB}{0, 0, 255}

\newcommand{\red}[1]{\textcolor{red}{#1}}

\begin{document}

\algnewcommand\algorithmicswitch{\textbf{switch}}
\algnewcommand\algorithmiccase{\textbf{case}}
\algnewcommand\algorithmicassert{\texttt{assert}}
\algnewcommand\Assert[1]{\State \algorithmicassert(#1)}%
% New "environments"
\algdef{SE}[SWITCH]{Switch}{EndSwitch}[1]{\algorithmicswitch\ #1\  }{\algorithmicend\ \algorithmicswitch}%
\algdef{SE}[CASE]{Case}{EndCase}[1]{\algorithmiccase\ #1}{\algorithmicend\ \algorithmiccase}%
\algtext*{EndSwitch}%
\algtext*{EndCase}%

\bstctlcite{IEEEexample:BSTcontrol}

\title{Optimal Switching Networks for Paired-Egress Bell State Analyzer Pools
\thanks{This work was supported by JST [Moonshot R\&D Program] Grant Numbers [JPMJMS226C] and [JPMJMS2061].}
}

\author{
\IEEEauthorblockN{
    Marii Koyama\IEEEauthorrefmark{5},
    Claire Yun \IEEEauthorrefmark{6},
    Amin Taherkhani\IEEEauthorrefmark{1},
    Naphan Benchasattabuse\IEEEauthorrefmark{1}\IEEEauthorrefmark{3},\\
    Bernard Ousmane Sane\IEEEauthorrefmark{1}\IEEEauthorrefmark{3},
    Michal Hajdu\v{s}ek\IEEEauthorrefmark{1}\IEEEauthorrefmark{3},
    Shota Nagayama\IEEEauthorrefmark{4}\IEEEauthorrefmark{1},
    Rodney Van Meter\IEEEauthorrefmark{5}\IEEEauthorrefmark{3}
}\\

\IEEEauthorblockA{\IEEEauthorrefmark{1}\textit{Graduate School of Media and Governance, Keio University Shonan Fujisawa Campus, Kanagawa, Japan}}
% \IEEEauthorblockA{\IEEEauthorrefmark{2}\textit{Graduate School of Science and Technology, Kanagawa, Japan}}
\IEEEauthorblockA{\IEEEauthorrefmark{3}\textit{Quantum Computing Center, Keio University, Kanagawa, Japan}}
\IEEEauthorblockA{\IEEEauthorrefmark{4}\textit{mercari R4D, Mercari, Inc., Tokyo, Japan}}
\IEEEauthorblockA{\IEEEauthorrefmark{5}\textit{Faculty of Environment and Information Studies, Keio University Shonan Fujisawa Campus, Kanagawa, Japan}}
\IEEEauthorblockA{\IEEEauthorrefmark{6}\textit{Department of Information Science, College of Agriculture and Life Science, Cornell University, New York, United States}}
\{mia, cly29, amin, whit3z, bernard, michal, rdv\}@sfc.wide.ad.jp, \{shota\}@qitf.org}

\maketitle
\thispagestyle{plain}
\pagestyle{plain}

\begin{abstract}
To scale quantum computers to useful levels, we must build networks of quantum computational nodes that can share entanglement for use in distributed forms of quantum algorithms. In one proposed architecture, node-to-node entanglement is created when nodes emit photons entangled with stationary memories, with the photons routed through a switched interconnect to a shared pool of Bell state analyzers (BSAs).  Designs that optimize switching circuits will reduce loss and crosstalk, raising entanglement rates and fidelity. We present optimal designs for switched interconnects constrained to planar layouts, appropriate for silicon waveguides and Mach-Zehnder interferometer (MZI) $2 \times 2$ switch points. The architectures for the optimal designs are scalable and algorithmically structured to pair any arbitrary inputs in a rearrangeable, non-blocking way. For pairing $N$ inputs, $N(N - 2)/4$ switches are required, which is less than half of number of switches required for full permutation switching networks. An efficient routing algorithm is also presented for each architecture. These designs can also be employed in reverse for entanglement generation using a shared pool of entangled paired photon sources.

\end{abstract}


\section{Introduction}

The importance of switching of signals has been understood since the earliest days of telecommunications.
The mid-twentieth century saw advances in both the practice and theory of switching for telephone networks, resulting in multi-stage designs such as Clos, Bene\u{s}, omega and butterfly networks for electrical signals, coupling small switch units together via discrete wires or coaxial cables~\cite{clos1953study,opferman1971class,benevs1965mathematical,benevs1964optimal, waksman1968permutation}. 
Inspired by these designs, and facing the need to scale up systems, computer architects have built \emph{multicomputers}, systems with many independent processors and memory units connected via interconnection networks~\cite{athas:multicomputer, benevs1965mathematical, dally04:_interconnects, awschalom2022roadmap}.
Multicomputer designs for quantum computers, in which a number of independent quantum computers with separate quantum registers and control systems are coupled via an interconnect network, are widely seen as a necessary architectural approach to achieving scalable, fault-tolerant systems~\cite{jiang07:PhysRevA.76.062323,jiang2010scalable,kim09:_integ_optic_ion_trap,kim05:_system,lim05:_repeat_until_success,PhysRevA.89.022317,nickerson2013topological,oi06:_dist-ion-trap-qec,van-meter06:thesis,van-meter16:_ieee-comp,yimsiriwattana2004distributed,diadamo21:dist-vqe,satoh2020federated}.
 For multicomputer quantum processing of transferring quantum state (teledata) or performing teleportation of quantum gates (telegate), the ability to generate Bell pairs and deliver them to arbitrary quantum processing units is essential~\cite{van2006distributed,van2007communication,caleffi2022distributed}.   

In addition to the architectural challenges of solving a large scale quantum algorithm using a quantum multicomputer, the cooperative nature of some distributed quantum algorithms over a structured quantum network requires an efficient interconnect between quantum nodes in the network.
To achieve scalability, the unrealistic, abstract model with direct interfaces between every pair of quantum computers or nodes in the network must be replaced with a realistic switch interconnect architecture \cite{awschalom2022roadmap,gauthier2023architecture} to have reconfigurable paths between arbitrary nodes.  
Switching interconnects are also indispensable components of a number of quantum network testbed designs that are planned to be deployed in the near future~\cite{alshowkan2012reconfigurable,bersin2024development}, paving the way to the eventual quantum Internet~\cite{kozlowski2023rfc, hajduvsek2023quantum, wehner2018quantum, vanmeter2022quantum, van2014quantum}.

\begin{figure*}[t]
   \centering
   \include{img/Sw}
   \includegraphics[width=0.95\textwidth]{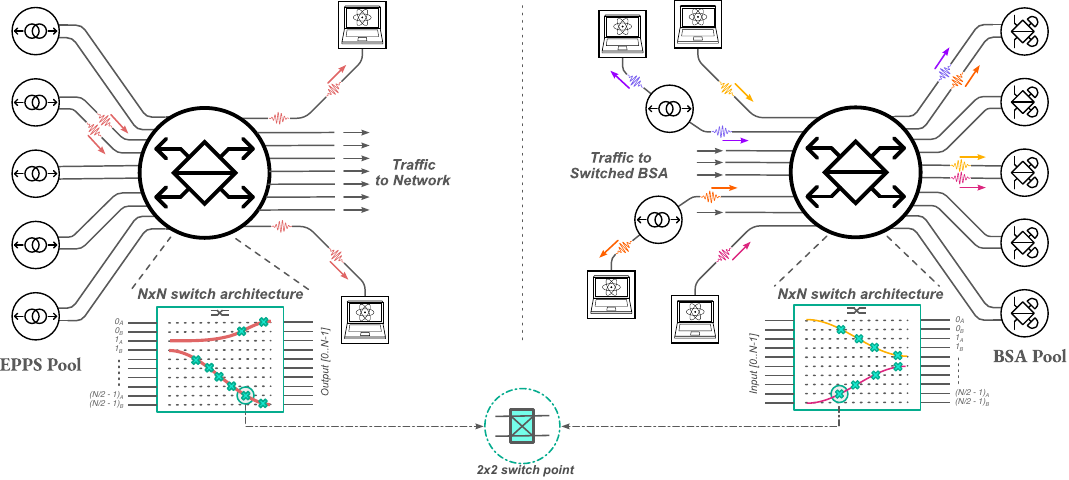}
   \caption{Deployment of switches in quantum networks for two possible scenarios: Left: Entanglement generation using shared EPPSs for two nodes in the network. Right: Entanglement consumption for entanglement swapping using shared BSAs. As in the lower portion of the figure, a switch can be composed of sets of $2\times 2$ switching points. Because distributing Bell pairs via EPPS (left) or performing entanglement swapping via BSA (right) are not location-dependent operations, we are free to select any reachable EPPS or BSA, offering the opportunity to reduce switch complexity compared to classical full permutation switches.}
   \label{fig:sw-highlevel-arch}
\end{figure*}

The primary service of quantum networks is the distribution of entangled states, usually entangled pairs of qubits.
This departure from the packet-forwarding or circuit-switching nature of classical networks presents a unique set of challenges for the design of quantum switching networks, with the goal of distributing pairs of entangled photons to the terminals. 

One approach to quantum switching interconnects assumes the availability of a shared pool of \textit{entangled photon-pair sources} (EPPS Pool) as pictured in the left panel of Fig.~\ref{fig:sw-highlevel-arch}.
In this architecture, pairs of entangled photons are generated at the EPPS nodes and routed by the optical switch to the appropriate end nodes.
This demonstrates the fundamental difference between classical and quantum switching interconnects.
In the EPPS Pool architecture, initially neighboring inputs (entangled photon pairs) must be switched to an arbitrary pair of outputs.
A non-planar optimal solution to this problem was proposed by Drost \emph{et al.}~\cite{drost2016switching} for cases up to 10$\times$10 quantum switches.
This solution was found via exhaustive search but did not provide a good recursive design that would lead to optimal and scalable quantum switching networks.

Optimal planar and scalable designs for full permutation $N\times N$ switching networks are over-designed for the problem of pair matching as they require the ability to route all $N!$ input-output combinations.
The ability to route photons to arbitrary BSAs, including arbitrary choice of the BSA ports, is not required for successful execution of entanglement swapping between desired pairs of photons.

We consider the inverse problem to the EPPS Pool architecture, shown in the right panel of Fig.~\ref{fig:sw-highlevel-arch}.
Photons originating from the quantum network are inputs into the quantum switch, which routes the desired pairs of photons to be incident onto the same \textit{Bell State Analyzer} (BSA).
The pair of photons then undergoes a Bell-state measurement, leading to entanglement between the respective end nodes.
The unique aspect of this BSA Pool architecture is that \emph{which BSA is used to perform entanglement swapping is irrelevant}.

We propose three recursive designs for a planar $N\times N$ quantum switch composed of a number of $2\times2$ switch points.
We obtain a lower bound for the number of switch points required for the quantum switch to be rearrangeably non-blocking, and demonstrate that all three of our designs saturate this lower bound.
For each design, we present an efficient routing algorithm.
We further analyze the depth of the three designs, and the average number of switch points that a photon traverses in order to better understand their loss properties.
Finally, we compare our planar designs with existing planar~\cite{spanke1987n} and non-planar solutions for quantum~\cite{drost2016switching} as well as classical switches~\cite{benevs1964optimal,waksman1968permutation}, and demonstrate favourable scaling properties of our designs.

\section{Preliminaries}

We begin by describing the basics of optical switching networks and how quantum networks differ from classical ones.
We then proceed to summarize fabrication factors that place constraints on our design.
Finally, we discuss the assumptions used in this work before ending this section with the problem statement.

\subsection{Classical and quantum optical switching networks} \label{subsec:OpticalSwitching}

An optical switching network has a number of important characteristics that should be considered while designing a switching configuration:

\begin{itemize}
\item {\bf Size:} the number of input and output ports
\item {\bf Blocking/non-blocking:} whether the network can handle all possible input/output combinations
\item {\bf Switching time:} the reconfiguration time for the network
\item {\bf Propagation delay:} the time needed for photons to cross the network
\item {\bf Insertion loss:} the probability of losing photons when crossing the physical interface from the channel to the switching element, typically involving a fiber/air boundary or a fiber/chip interface (generally reported in dB)
\item {\bf Switching loss:} the probability of losing photons within the switching element (generally reported in dB)
\item {\bf Crosstalk:} the leakage of signal to undesired transmission paths (generally reported in dB)
\item {\bf Redundancy:} whether or not connectivity is degraded if a switch point or link fails
\item {\bf Physical dimensions:} especially important when considering integration into the system
\item {\bf Cost:} often reported as per-port cost for large configurations
\end{itemize}

The above list is largely common between classical and quantum switching networks.  In a classical network, within reason, loss can be compensated for by increasing input signal power.
In the systems we are designing (Fig.~\ref{fig:sw-highlevel-arch}), the networks instead carry single photons.

Loss becomes the most critical metric and increases as the photons pass through channels, interfaces and switch points.

Polarization may be critical in quantum networks (depending on choice of qubit representation), and it can be changed by each component.  If the network itself can be built with enough stability that its effect on polarization can be characterized at infrequent intervals, network operation will be more efficient~\cite{krutyanskyi2023entanglement,crespi2011integrated,corrielli2014rotated}. Moreover, in physical design, self-calibration can also be achieved by adding auxiliary optical components to the photonic core~\cite{xu2022self,bogaerts2020programmable}.

Most of the above characteristics depend on the choice of physical fabrication technology, but from the architectural design point of view, the number of input ports, total number of switches, and circuit depth indirectly affect propagation delay, insertion loss, switching loss, crosstalk, physical dimensions and cost.
\subsection{Fabrication Considerations} \label{subsec:FabricationReq}

For optical systems, we can build switching systems based on photonic integrated circuit (PIC) technology~\cite{wang2020integrated,bogaerts2020programmable}, or via free-space propagation of light, e.g. using micro electro-mechanical system (MEMS) switches~\cite{saleh2019fundamentals,kim03:_1100_port_mems}.
PICs have relatively high loss per centimeter of waveguide, but have the advantage of fewer fiber/air or fiber/chip interfaces that must be crossed compared to designs that use discrete components for each switch point. 
In this paper, we focus on integrated circuits.

Photolithographic fabrication for waveguides~\cite{harris2016large} or photonic crystals~\cite{lonvcar2000waveguiding, o2009photonic} results in planar layouts.
Two waveguides can be run close to each other, resulting in $2\times 2$ switch units based on Mach-Zehnder interferometric techniques, allowing two photons to be routed straight through or swapped under programmatic control.
Grids of these basic units take sets of inputs from one edge of a chip and route them to a set of outputs at the opposite edge of the chip.

Promising platforms for fabrication of integrated photonic circuit such as silicon on insulator~\cite{harris2016large}, silica on silicon~\cite{o2009photonic} and stoichiometric silicon nitride  (\chem{Si_3N_4})~\cite{taballione20198} have different fabrication complexity, photon loss and index contrast with different supported wavelengths.
These layouts are generally kept planar due to the difficulty of
routing a waveguide off the substrate surface without significant
loss~\cite{nesic2019photonic}.
Therefore, we focus exclusively on planar switch designs in this work.

\subsection{Assumptions}

Following the discussion in Sec.~\ref{subsec:OpticalSwitching} and Sec.~\ref{subsec:FabricationReq}, we now outline the assumptions used in the rest of the paper.
The switch waveguides carry individual photons generated from quantum memories or entangled photon pair sources.
BSAs are located outside the switch, with each BSA being attached to two neighboring output ports of the switch.
Photons are to be matched in pairs and routed to any available BSA (\emph{paired egress}).
Use of the switch occurs in independent rounds, between which the switch may be reconfigured.
The design must be planar and non-blocking for all possible choices of input pairings.
Switch points are based on $2 \times 2$ switching elements compatible with basic building blocks in photonic integrated circuits.
Photons are guided single-pass only, from one side of the switch to another, and without any recirculation.

Issues of achieving indistinguishability between the input photons, such as polarization maintenance, spectral properties and time of arrival of the photons at the BSA, are out of the scope for this work.
Based on the above assumptions, we can define the problem to be addressed.

\subsection{Problem statement} 

Consider an $N\times N$ switching network with $N$ input ports, with photons $X_0,X_1,\ldots X_{N-1}$ where photon $X_i$ comes in at the $i$-th port, and $N$ output ports coupled to $N/2$ BSAs.
A paired egress request is represented as a tuple $(X_i,X_j)$, where $0 \leq i < j < N$, and we call the set of all required pairings $PL_{N/2}=\{(X_i,X_j)\}$, where $i$ and $j$ each appear exactly once.

We want to find a scalable, planar and rearrangeably non-blocking topology for a switching network composed of base $2\times 2$ switch points, and a corresponding efficient routing algorithm capable of handling an arbitrary pair list $PL_{N/2}$.
The routing algorithm must provide the state of every switch point in the network, denoted by $SW^l_i\in\{\sbar,\scross\}$, where $SW^l_i$ represent the switch at layer $l$ between line $i$ and $i+1$,

and a permuted list of photons where the two photons at index $2j$ and $2j+1$ of the list go into the same BSA (the $BSA_j$) after exiting the switch.
The concept of a layer will be made more precise when we introduce our designs.

\if0
\begin{table}[t]
    \centering
    \caption{Terminology \rdv{If every variable is introduced properly in the paper, maybe we don't need this any more?}}
    \label{tab:hogehoge}
    \begin{tabular}{ll}
        \hline \hline
        Input     &   Indexed from $X_0$ to $X_{N-1}$ from top to bottom.\\
        BSA		    & 	Indexed from \sw{BSA}{0} to \sw{BSA}{N/2-1} from top to bottom.\\
        
        $(X_i,X_j)$		&	Paired inputs.\\
        $SW$ & Total number of switch points across all layers. \\
        $SW^k$  & Number of switch points in layer $k$. \\
        $SW^{k}_{i}$  & Switch point at layer $k$ between line $i$ and $i+1$. \\
       
        $PL_{K}$ & An input paired list for matching at $K$ BSAs. \\  
        & \red{need referencing to the table in text.} \\
        
        \hline
    \end{tabular}
\end{table}
\fi

\begin{figure*}[t]
  \centering
  \includegraphics[width=0.75\textwidth]{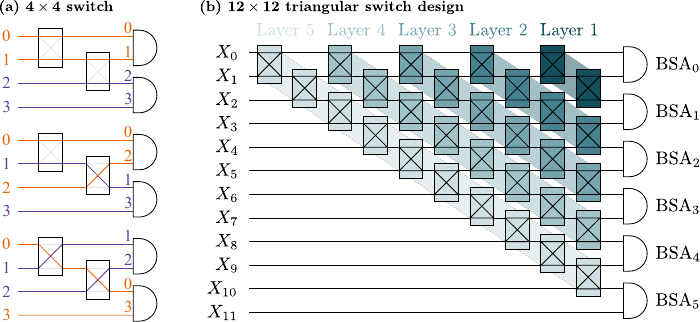}
  \caption{(a) 4$\times$4 non-blocking planar paired-egress switching interconnect. Changing the state of the 2$\times$2 switch points achieves all possible photon pairings.
  (b) A rearrangeable non-blocking $12\times 12$ planar paired-egress switching interconnect.
  Switch points of the same signify that they are added within the same network size and constitute a \emph{layer}. Semicircles represent BSAs.}
  \label{fig:triangular}
\end{figure*}

\section{Optimal number of switch points}
\label{sec:lowerbound}

In order to show that our proposed switching networks are optimal, we consider the minimum number of switch points required to pair all $N$ input photons.
It is known that a classical planar $N\times N$ switching network requires at least $N(N-1)/2$ switch points to achieve all input-output permutations~\cite{spanke1987n}.

We adapt the techniques used in~\cite{spanke1987n} for the case of paired-egress switching networks.
In order to obtain the minimum number of switch points for the network to be rearrangeably non-blocking, we consider the worst case scenario, where always the two most distant input photons are required to be paired together.
The list of pairings is therefore given by $(X_0,X_{N-1})$, $(X_1,X_{N-2})$, $(X_2,X_{N-3})$, and so on.
Bringing the input photons $X_0$ and $X_{N-1}$ together means that we must swap with all the other $N-2$ input photons, requiring $N-2$ switch points.
This is true regardless of which BSA is assigned to perform entanglement swapping on the input photons $X_0$ and $X_{N-1}$.
A sample path and its required swaps are shown in the lower portion of the right panel of Fig. \ref{fig:sw-highlevel-arch}.
Bringing the next pair of photons, $(X_1,X_{N-2})$, together requires $N-4$ swaps.
Continuing with this logic, the minimum number of swaps, and therefore switch points, is given by
\begin{equation}
    \sum_{k=1}^{N/2-1} (N -2k) = N(N-2)/4.
    \label{eq:minimum_bound}
\end{equation}
We observe that the minimum number of switch points for a planar $N\times N$ rearrangeably non-blocking network with paired egresses is less than half of that obtained in~\cite{spanke1987n} for a classical switching network.

\section{Triangular Design} \label{sec:triangular}

We begin with the simplest planar configuration for a rearrangeable non-blocking switching network with the minimum number of switch points.  From this point forward in the paper, all designs are assumed to be paired egress, so we dispense with the qualifier.

\subsection{Architecture}

The smallest non-trivial switching network is a $4\times4$ switch, shown in Fig.~\ref{fig:triangular}(a), and requires at least 2 switch points.
By changing the state of the switch points, it is possible to achieve all possible input photon pairings.
For example, leaving both switch points in the \sbar state results in entanglement swapping between input photon pairs (0,1) and (2,3).
Turning both of the switch points to \scross state routes the input photons in such a way that entanglement swapping is performed on the pairs (1,2) and (0,3).

This $4\times4$ switch forms the basic building block for the triangular switch architecture, as depicted in Fig.~\ref{fig:triangular}(b).
For $N$ input photons, the switch is composed of $N/2-1$ layers, where layer $k$ contains $SW^k=2k$ switch points.
The total number of switch points is therefore given by
\begin{equation}
    SW_{\text{triangular}} = \sum_{k=1}^{N/2-1} SW^k = N(N-2)/4.
\end{equation}
We see that the triangular architecture is optimal in terms of the number of switch points.
The switch is constructed recursively by adding all switch points within a layer in a cascaded fashion.
For layer $k$, the switch points are arranged in the following way,
\begin{equation}
    SW^k_0 \rightarrow SW^k_1 \rightarrow \ldots \rightarrow SW^k_{2k-1},
\end{equation}
where $SW^l_i$ represent the switch at layer $l$ between line $i$ and $i+1$, as shown in Fig.~\ref{fig:triangular}(b) for the case of $N = 12$.

\subsection{Routing}

Showing that a switch design is rearrangeably non-blocking amounts to demonstrating that given a pair list $PL_{N/2}$, it is always possible to find a configuration of all switch points that permutes the input photon list such that all photon pairs are adjacent to each other.
The routing algorithm strongly depends on the design of the switch.
For the triangular design, it is relatively straightforward and is presented in Algorithm~\ref{alg:triangular-while-loop}.

The input to the routing algorithm is given by the input photon list $(X_0,\ldots,X_{N-1})$, and the pair list $PL_{N/2}$.
The main strategy, similar to the bubble sort algorithm, is to start with photon $X_{N-1}$ since it is not incident onto any switch points and cannot be routed.
We find its partner $X_j$, such that $(X_j,X_{N-1}) \in PL_{N/2}$, and configure the switch points in layer $N/2-1$ in order for the pair to meet on lines $(N-2, N-1)$.
This is achieved by setting all switch points $SW^{N/2-1}_k$, for $j\leq k<N-2$ to the \scross state.
All other switch points in the layer, $SW^{N/2-1}_k$ where $0\leq k<j$, are set to the \sbar state.
The effect of this configuration leaves the ordering of photons $X_0,\ldots,X_{j-1}$ unaffected, cascades the photon $X_j$ down to line $N-2$, and shifts all remaining photons up by one unit.
This process is next repeated for the next layer with input size $N-2$, and the new list of photons $(X_0,\ldots,X_{j-1}, X_{j+1},\ldots, X_{N-2}, X_{j}, X_{N-1})$ until the input size becomes 2, reaching the trivial case.

For each layer, we traverse the photon list of length $2l+2$ once and assign the switch states of $2l$ switches.
Therefore, the time complexity of this routing algorithm is $O(N^{2})$.

%\

\newenvironment{algocolor}{%
   \setlength{\parindent}{0pt}
   \itshape
   \color{purple}
}{}
\begin{algorithm}
    \caption{Routing algorithm for the triangular design}
    \label{alg:triangular-while-loop}
\hspace*{\algorithmicindent}\textbf{Input:} $Photons$: indexable list $(X_0, X_1, X_2, \ldots, X_{N-1})$, \\
\hspace*{\algorithmicindent} \hspace*{0.8cm} $PL$: set be paired $\{(X_i, X_j)  \mid 0 \leq i, j \leq N-1\}$ \\
\hspace*{\algorithmicindent}\textbf{Output:} permuted list: $\pi((X_0, X_1, \ldots, X_{N-1}))$ \\
\hspace*{\algorithmicindent} \hspace*{0.9cm} Set of switch states $\{ SW_j^l \}$ 
\begin{algorithmic}[1]
\Procedure{Routing\_Triangle}{}($Photons$, $PL$)
\State $n \gets$ length of $Photons$ \Comment{initially $n = N$}
\State $SW \gets \emptyset$
\While{$n > 2$}
    \State $l \gets n/2-1$ \Comment{Current switch layer}
    \State $i \gets$ index of $Photons[n-1]$'s partner
    \State $SW \gets SW + \{SW_j^l\ = Bar \mid 0 \leq j < i\}$
    \State $SW \gets SW + \{SW_j^l\ = Cross \mid i \leq j \leq n-3\}$
    \State $Photons \gets$ move $Photons[i]$ to pos. $n-2$
    \State $n \gets n - 2$
\EndWhile
\State \Return $Photons$, $SW$
\EndProcedure
\end{algorithmic}
\end{algorithm}

\section{Chevron Design} \label{sec:Chevron}

Our second switch design redistributes the switch points in a more uniform manner.
This change leads to a more complicated, yet still efficient, routing algorithm.

\subsection{Architecture} \label{subsec:Chevron-Arch}

\begin{figure}[t]
  \centering
  \includegraphics[width=\columnwidth]{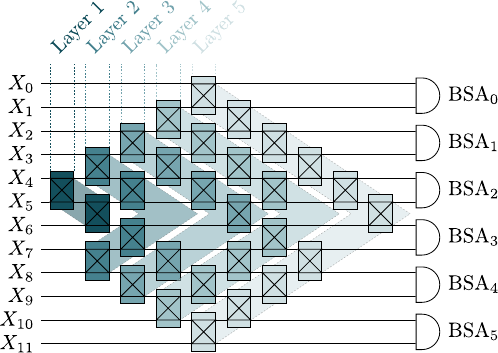}
  \caption{A chevron shaped 12 $\times$ 12 non-blocking planar paired-egress switching design. For an $N \times N$ network with odd $N/2$, a chevron shaped set of switch points, without switch point on line $N/2-1$ and $N/2$, is added. The total number of switch points is $N(N-2)/4$.}
  \label{fig:chevron}
\end{figure}

The chevron design is pictured in Fig.~\ref{fig:chevron} for the case of $N=12$.
Similar to the triangular design, the chevron design consists of $N/2-1$ layers with layer $k$ consisting of $SW^k=2k$ switch points, resulting in the optimal number of total switch points $N(N-2)/4$.
However, the switch points are laid out in a chevron, rather than a single diagonal arrangement.

For layer $l$, with $l$ being even, half of the switch points in that layer are placed in the upper half of the chevron, with the rest of the switch points being placed in the bottom half.
More precisely, the switch points in the upper half of a layer are placed as follows, 
\begin{align}
    SW^l_{N/2-l-1} \rightarrow SW^l_{N/2-l} \rightarrow \ldots \rightarrow SW^l_{N/2-2}.
\end{align}
The switch points in the bottom half of the layer mirror the above arrangement,
\begin{align}
    SW^l_{N/2+l-1} \rightarrow SW^l_{N/2+l-2} \rightarrow \ldots \rightarrow SW^l_{N/2}.
\end{align}
For odd layers, the placement of the switch points is similar with the exception that the switch point $SW^l_{N/2}$ is removed and a new switch point $SW^l_{N/2-1}$ is placed at the tip of the chevron, as shown in Fig.~\ref{fig:chevron}. 

\subsection{Routing}

Algorithm~\ref{alg:chevron-shorter} illustrates the procedure for determining switch status and the permuted list in the routing process.
The routing algorithm for the chevron operates on the principle that if all photon pairs entering the last layer are adjacent to each other, except for two adjacent photons requiring pairing with the top- and bottom-most photons, the last chevron layer can handle their pairing.
Alternatively, if all photons are already paired except for the top- and bottom-most photons, the algorithm can address this scenario as well. 
This approach is valid as we recursively go from layer $N/2-1$ to $N/2-2$ reducing the size of the switch from $N \times N$ to $(N-2) \times (N-2)$ by virtually pairing the photons that need to be matched with the top- and bottom-most photons until reaching the trivial inputs $2 \times 2$.
This guarantees that at every recursion, all the photons will have a pairing within the considered range, either true or virtual pairings.

\begin{figure*}[t]
  \centering
  \includegraphics[width=0.8\textwidth]{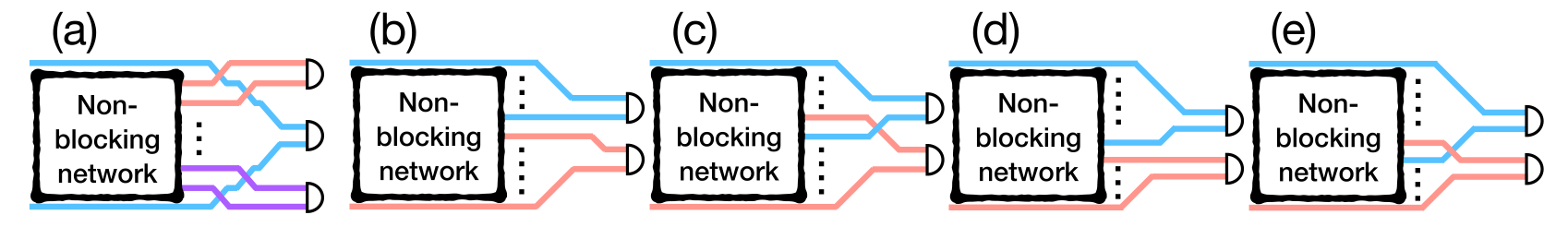}
  \caption{Recursively constructing larger chevron circuits (Sec.~\ref{sec:Chevron} and Fig.~\ref{fig:chevron}) requires handling five possible cases as two inputs are added, one at the top and one at the bottom.
  The same colored lines denote the location of photons to be paired. 
  (a) represents the case pairing the top-most photon with the bottom-most photon. In this case all the other input ports are paired using previously-constructed internal non-blocking network. After applying the last chevron layer (not shown) the paired photons shift up or shift down to finally pair in the BSAs layer. 
  (b), (c) represents the case that the photons at the top-most port and bottom-most port have to be paired with two other photons in upper half part of the switch configuration.
  In (b) the two pairs face each other. 
  In (c) the two pairs face the opposite side.  
  (d), (e) represents the case that the photons at the most top port and most bottom port have to be paired with two other photons in lower half part of the switch configuration.
  In (d) the two pairs face each other.
  In (e) the two pairs face the opposite side.  
} 
  \label{fig:fourcases}
\end{figure*}

Determining where the pairs meet and which switch states to set at layer $l$ involves examining two different scenarios. 
First, consider the case where the top- and bottom-most photons form a pair, while the other photon pairs (either true or virtual) are already adjacent.
In this scenario, all switch points in layer $l$ are set to \scross{}, causing the two outer photons to meet at the middle ($N/2$, $N/2+1$) or ($N/2-1$, $N/2$)
when $l$ is odd or even, respectively.
However, if the top- and bottom-most photons need to pair with two virtual pairings from the previous layer (on lines $i$ and $i+1$), and the virtual pair is oriented incorrectly, the switch point $SW_{i}^l$ is set to \scross{}.
Consequently, the two outer photons adjust to meet their partners, and if the pair resides in a different half (e.g., the top-most's partner is in the bottom half), its partner also moves to meet the outer qubit at the middle.
This results in only one switch point, aside from possibly $SW_{i}^l$, being set to \sbar{}; $SW_i^l$ when 
or $SW_{i+1}^l$ otherwise, while the remaining switch points are set to \scross{}.
All the possible scenarios and how the photons are moved are depicted in Fig.~\ref{fig:fourcases}.

The runtime of the routing algorithm for the chevron architecture is $O(N^2)$.
In each recursion, we go through the photon list only once to find partners of the top- and the bottom-most photons and all the switch states in the layer can then be decided just by knowing where the top- and bottom-most photons need to be moved to.

\begin{algorithm*}[t]
    \caption{Routing algorithm for chevron design}
    \label{alg:chevron-shorter}
\hspace*{\algorithmicindent}\textbf{Input:} $Photons$: indexable list $(X_0, X_1, X_2, \ldots, X_{N-1})$, \\
\hspace*{\algorithmicindent} \hspace*{1cm} $PL$: set of photons to be paired $\{(X_i, X_j)  \mid 0 \leq i, j \leq N-1\}$, \\
\hspace*{\algorithmicindent} \hspace*{1cm} $SW$: set of switch states (initially null set), \\
\hspace*{\algorithmicindent} \hspace*{1cm} $left$: indicating the start switch line considered in the recursion \\
\hspace*{\algorithmicindent}\textbf{Output:} permuted list: $\pi((X_0, X_1, \ldots, X_{N-1}))$, \\
\hspace*{\algorithmicindent} \hspace*{1.3cm} Set of switch states $\{ SW_k^l \}$ 
\begin{algorithmic}[1]
\Procedure{Routing\_Chevron}{}($Photons$, $PL$, $SW = \emptyset$, $left = 0$)
\State $n \gets$ length of $Photons$ \Comment{first call to the procedure will have $n = N$}
\If{$n$ is equal to 2}
    \State \Return $Photons$, $SW$
\EndIf
\State $l \gets n/2-1$ \Comment{Current switch layer}
\State $top, bot \gets Photons[0], Photons[n-1]$
\State $Photons \gets$ remove $top$, $bot$ from $Photons$

\If{$(top, bot) \in PL$}
   
    \If{n/2 is even}
        \State $i \gets left + n/2$    
    \Else
        \State $i \gets left + n/2 - 1$    
    \EndIf
    \State $SW \gets SW + \{SW_j^l\ = Cross \mid \text{for all } j\}$
    \State $Photons$, $SW \gets$ \Call{Routing\_Chevron}{$Photons$, $PL$, $SW$, $left+1$} 
    \State $Photons \gets$ insert $top, bot$ into $Photons$ at position $i$, $i+1$
    \State \Return $Photons$, $SW$
\EndIf

\State $top'$, $bot' \gets$ partner photons of  $top$ and $bot$ respectively
\State $PL \gets PL - \{(top, top'), (bot, bot')\} + \{(top', bot')\}$ \Comment{virtual pair}
\State $Photons$, $SW \gets$ \Call{Routing\_Chevron}{$Photons$, $PL$, $SW$, $left+1$} 
\State $i, j \gets$ indices such that $Photons[i] = top'$, $Photons[j] = bot'$
\If{$i < j$}
    \State $SW \gets SW + \{SW^l_{left + i} = Bar\}$
\Else
    \State $i, j \gets j, i$
    \State $Photons[i], Photons[j] \gets top', bottom'$
\EndIf
\If{ $i$ is in top half}
    \State $SW \gets SW + \{SW^l_{left + i - 1} = Bar\}$
    \State $Photons \gets$ insert $top$ on the left of $top'$
    \State $Photons \gets$ remove $bot'$ and insert $bot', bot$ at the middle
\Else
    \State $SW \gets SW + \{SW^l_{left + j} = Bar\}$
    \State $Photons \gets$ insert $bot$ in on the right of $bot'$
    \State $Photons \gets$ remove $top'$ and insert $top, top'$ at the middle
\EndIf
\State $SW \gets SW + \{SW^l_k = Cross \mid \text{for other unset switches in the layer} \}$ 
\State \Return $Photons$, $SW$
\EndProcedure
\end{algorithmic}
\end{algorithm*}

\section{Brickwork Design} \label{Brikwork Design}

%Strcutrure and layout:
\begin{figure}[t]
  \centering
  \includegraphics[width=0.95\columnwidth]{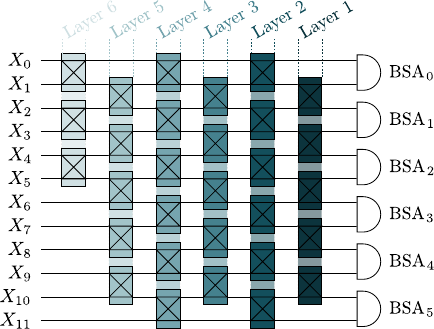}
  \caption{Brickwork design for a $12\times12$ switching network.}
  \label{fig:DesignIII}
\end{figure}

Our final design rearranges the switch points further into a brickwork pattern as shown in Fig.~\ref{fig:DesignIII}.

\subsection{Architecture}

This time, an $N\times N$ switch consists of $N/2$ layers.
There are three types of layers; odd, even, and the last layer.
Unlike previous designs, the number of switch points inside a layer is independent of the layer number $k$.

Odd layers contain $N/2-1$ switch points, while even layers contain $N/2$ switch points.
The brickwork design is given by the following switch point placement,
\begin{align}
\begin{split}
    SW^k_{1} & \rightarrow SW^k_{3} \rightarrow \ldots \rightarrow SW^k_{N-3}; \quad\text{for odd } k, \\
    SW^k_{0} & \rightarrow SW^k_{2} \rightarrow \ldots \rightarrow SW^k_{N-2}; \quad\text{for even } k.
    \label{eq:rules-brickwork}
\end{split}
\end{align}
The exception to this rule is the final layer that contains $\lfloor N/4 \rfloor$ switch points that are placed according to~\eqref{eq:rules-brickwork}.
The total number of switch points is therefore given by
\begin{equation}
    SW = \sum_{\text{odd }k} SW^k + \sum_{\text{even }k} SW^k = N(N-2)/4,
\end{equation}
which shows that the brickwork design is also optimal in terms of the number of switch points.

The brickwork design is similar to a full mesh structure of Spanke and Bene{\v{s}} in~\cite{spanke1987n}, with one difference.
The full mesh architecture requires $N$ layers to perform $N!$ full permutations, while in the brickwork design $N/2$ layers are sufficient for pairing an arbitrary input set.

\subsection{Routing}

The routing algorithm for pairing all the input photons in the brickwork design is shown in Algorithm~\ref{alg:Brickwork_routing}.
The underlying concept shares similarities with the Triangular design, but the process of removing switches is more intricate.
First, we identify the partner photon $X_{i}$ of photon $X_{N-1}$ and progressively shift it downward to meet with $X_{N-1}$.
If the lowest line $X_{i}$ can reach is line $j$, where $j \neq N-2$, we also shift photon $X_{N-1}$ up to line $j+1$.
This method ensures that upon removing all utilized switch points and wire segments traversed by $X_i$ and $X_{N-1}$, the resultant structure maintains a brickwork architecture, potentially with additional switch points in the last and second-to-last layers if the photon $X_i$ is shifted with the earliest switch points it encounters while $X_{N-1}$ is shifted as late as possible.
These extra switch points can be configured to the \sbar state, allowing for further recursion of the same routing algorithm until we reach the trivial photon parings.

\begin{algorithm*}[t]
\hspace*{\algorithmicindent} \textbf{Inputs:} \hspace*{0.1cm} $Photons$: indexable list $(X_0, X_1, X_2, \ldots, X_{N-1})$, \\ 
\hspace*{\algorithmicindent} \hspace*{1.3cm} $PL$: set of tuple photons to be paired $\{(X_i,X_j) \mid 0 \leq i,j \leq N - 1\}$,\\ 
\hspace*{\algorithmicindent} \hspace*{1.3cm} $SW$: set of switch states (initially null set)\\ 
\hspace*{\algorithmicindent} \textbf{Outputs:} Permuted list: $\pi((X_0, X_1, \ldots, X_{N-1}))$, \\
\hspace*{\algorithmicindent} \hspace*{1.3cm}
Set of switch states $SW^l_k$

\begin{algorithmic}[1]
\Procedure{Routing\_Brickwork}{$Photons$,  $PL_{N/2}$, $SW$}
\If{($N$ is equal to 2)}
    \State \Return $Photons$, $SW$
\Else
    \State Find $X_{i}$ $\ni (X_{i}, X_{N-1}) \in PL$ 
 
    \State Let $SW^{layer}_{j}$ be the first switch point $X_i$ encounters  \Comment{$(j=i$ or $i-1)$ and $(layer = N/2 -1$ or $N/2)$}
    \If{$j$ is equal to $i$}
        \State  $SW \gets SW + \{SW^{layer}_j\ =$ \scross$\}$ 
        \State $SW^{layer}_{j} = $ \scross
    \Else 
        \State  $SW \gets SW + \{SW^{layer}_j\ =$ \sbar$\}$ 
    \EndIf
    
    \While{$j < N-2$ or $layer > 0$} \Comment{create a diagonal path from $X_i$ toward $X_{N-1}$}
        \State $j$++, $layer${\tt --}
        \State  $SW \gets SW + \{SW^{layer}_j\ =$ \scross$\}$         
    \EndWhile

    \If{$j$ is equal to $N-2$}   \Comment{$X_i$ and $X_{N-1}$ are adjacent}
        \State  $SW \gets SW + \{SW^{L}_{N-2}\ =$ \sbar$ \forall L = 2l$ (even layers), $0 \leq 2l \leq N/2 - 1 \}$ 
        \State  $SW \gets SW + \{SW^{L}_{N-3}\ =$ \sbar$ \forall L = 2l + 1$ (odd layers), $1 \leq 2l + 1 < layer \}$
        \State $Photons[i]\gets N-2$ and $Photons[N-1] \gets N-1$
    \Else \Comment{$X_{N-1}$ needs to move up to be adjacent of $j$ in $j+1$}
    \State $Photons[i]\gets j$ and $Photons[N-1] \gets j+1$
    \State $j$++
    \While{$j < N-2$} \Comment{A diagonal path is created to route $X_{N-1}$ to output port $j+1$}
        \State  $SW \gets SW + \{SW^{layer}_j\ =$ \scross$\}$ 
        \State $j$++ and $layer$++
    \EndWhile

    \State  $SW \gets SW + \{SW^{L}_{N-2}\ =$ \sbar$ \forall L = 2l,$ (even layers) $layer < 2l \leq  N/2 - 1\}$
    \EndIf

    \While{Number of Assigned Switches $< N - 2$} \Comment{Set other unused $SW$ on top of input port of $X_i$ to \sbar}
        \State  $SW \gets SW + \{SW^{layer = N/2 }_{k} =$ \sbar\}$ (0 \leq k < i$ and start with $layer = N/2$ then $N/2-1$)  
    \EndWhile

    \State $PL \gets PL - (X_i, X_{N-1})$
    \State $Photons^{\prime} \gets $ reindex $Photons$ without $i$ and $N-1$ 
    \State $Photons^\prime, SW \gets$\Call{Routing\_Brickwork}{$Photons^\prime$, $PL$, $SW$} 
    \State $Photons \gets merge(Photons^\prime,i, N-1, Photons[i], Photons[N-1])$  \Comment{put $X_i$ and $X_{N-1}$ in the final places}   
    \State \Return $Photons, SW$ 
    \EndIf
\EndProcedure
\caption{Routing Algorithm for the Brickwork design}
\label{alg:Brickwork_routing}
\end{algorithmic}
\end{algorithm*}

\begin{figure}[t]
    \centering
    \includegraphics[width=0.45\textwidth]{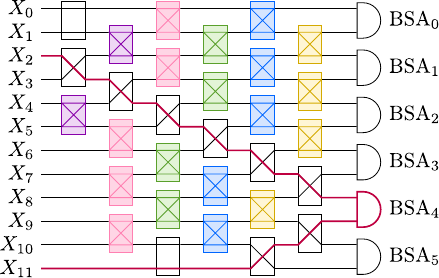}
    \caption{A pairing sample to show the behavior of the routing algorithm for a 12 $\times$ 12 brickwork structure. After pairing $X_{11}$ and its partner $X_2$ at $BSA_4$, we can conceptually redraw the switch with the committed path removed. After redrawing, the colored switch points will create a new (virtual) $10\times 10$ layering arrangement for the remaining switch configuration. All of the yellow switches will form the new first layer, all of the blue switches the new second layer, etc.}
    \label{fig:Routing_Sample_DesignIII}
\end{figure}

A sample routing round for pairing two photons ($X_{11}$ with $X_2$) in a $12 \times 12$ brickwork switch is shown in Fig.~\ref{fig:Routing_Sample_DesignIII}.
Given that the distance between $X_2$ and $X_{11}$ exceeds $N/2 + 1$ making it impossible for the pair to meet on line $(10, 11)$, requiring that both photons must move to pair up on the closest line to $X_{11}$, which is $(8, 9)$.
By applying \textit{move as soon as possible routing policy} for $X_2$ and \textit{move as late as possible routing policy} for $X_{11}$ results in the desired permuted list $(X_0,X_1,X_3,..,X_8,X_2,X_{11},X_9,X_{10})$.
We have now assigned \emph{nine} switch states, leaving the switch points into a brickwork design of smaller size with $SW_0^6$ being an extra switch. Therefore, as shown in top-left side of Fig.~\ref{fig:Routing_Sample_DesignIII}, we set it to the \sbar state.
These two move policies guarantee that after pairing the last unpaired photon with its partner will leave the switch in the Brickwork design and thus allowing the smaller Brickwork to be routed the same way.

Similar to routing algorithms in triangular and chevron designs, we traverse the photon list once per iteration, thus the runtime of the routing algorithm in brickwork design is also $O(N^{2})$.   

\section{Discussion}
\label{sec:discussion}

Having demonstrated that our three proposed planar designs are optimal and rearrangeable non-blocking, we now turn to a more detailed analysis of their depth properties.
The depth is directly related to the expected losses of the switching network.
We are not aware of any work considering planar designs for paired-egress outputs, making direct comparison impossible.
To place our work into the larger context of other switching network designs, we consider existing planar and non-planar designs, as well as classical permutation networks and shared EPPS pool networks.

\subsection{Switching network depth}

We first focus on the maximum depth, which quantifies the maximum number of switch points that a photon has to traverse.
For the Triangular design in Sec.~\ref{sec:triangular}, the maximum depth is $N-2$.
For the Chevron design in Sec.~\ref{sec:Chevron}, the maximum depth is $N-2$ when $N/2$ is even, $N-3$ when $N/2$ is odd.
The Brickwork design in Sec.~\ref{Brikwork Design} reduces the maximum depth to $N/2$.
The maximum depth of the Triangular and Chevron designs is close to the optimal design for classical planar $N\times N$ permutation networks, shown to be $N-1$ in~\cite{spanke1987n}.
The brickwork design on the other hand requires far fewer switch point traversals in the worst case scenario, reducing the overall loss in the switch.

\begin{figure}[htbp]
    \centering
    \includegraphics[width=0.5\textwidth]{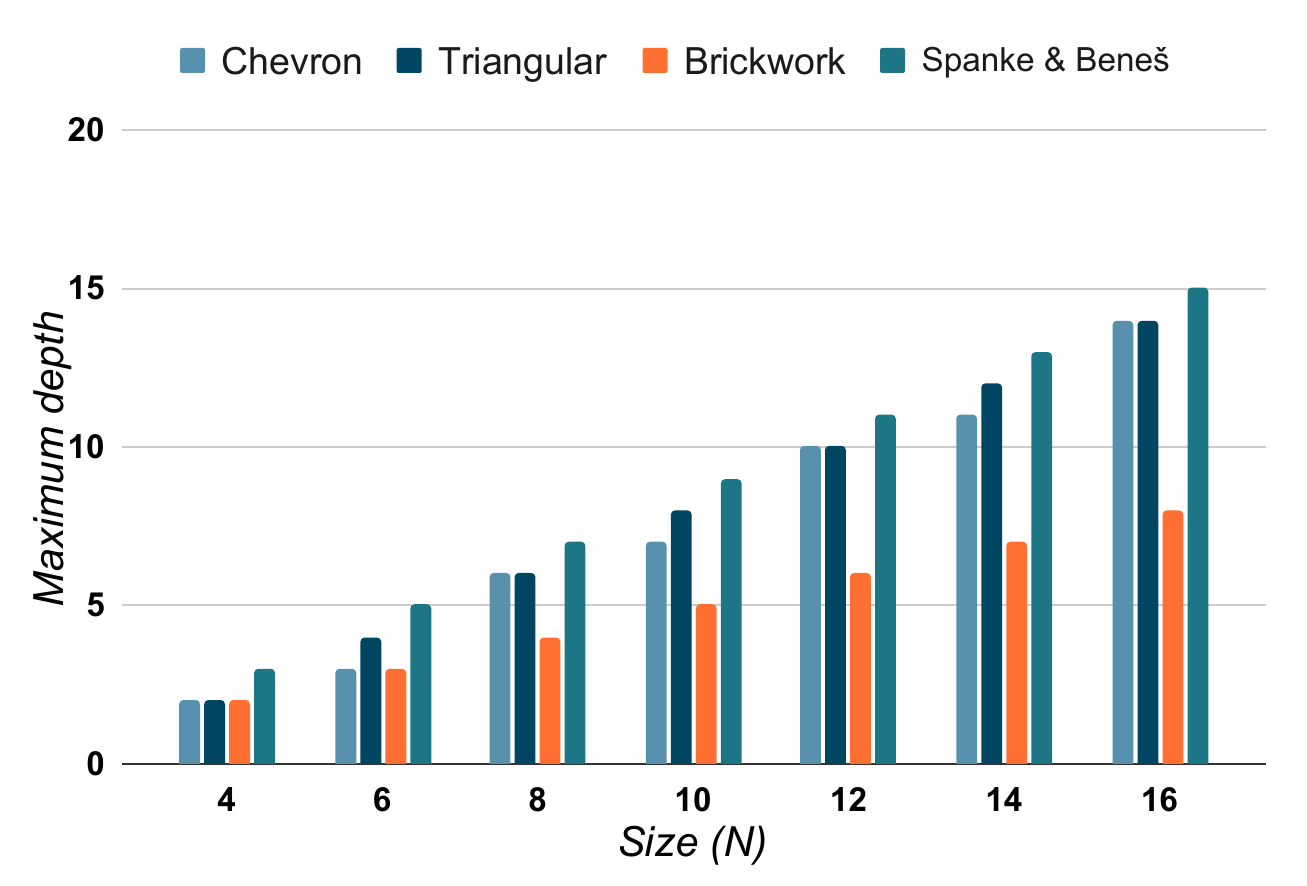}
    \caption{ Maximum depth required for pairing any combination of input $N$ in different scalable planar designs compared to the optimal full permutation planar design proposed by Spanke and Bene\u{s}~\cite{spanke1987n}.}
    \label{fig:depth}
\end{figure}

\begin{figure}[htbp]
    \centering
    \includegraphics[width=0.95\columnwidth]{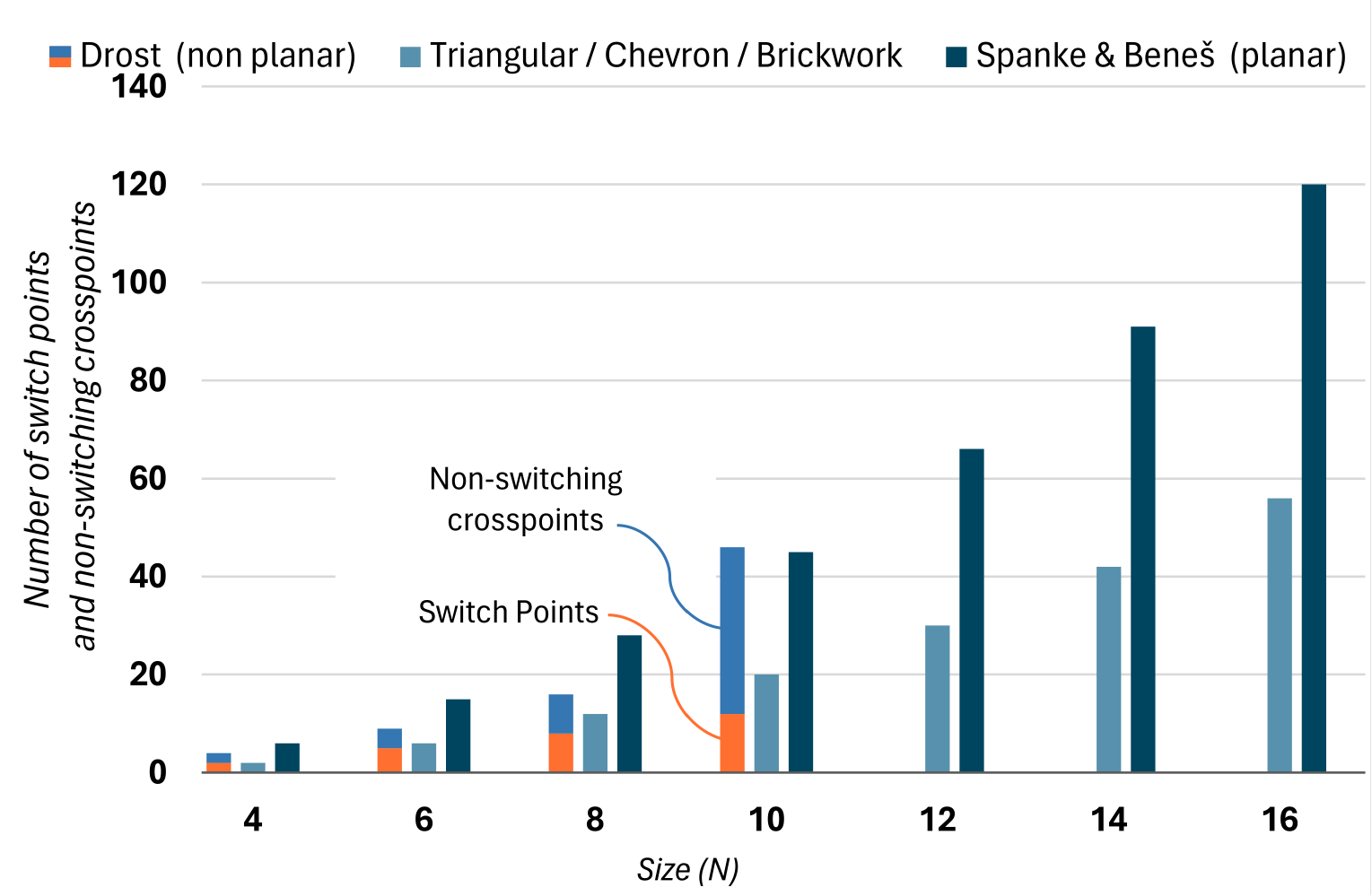}
    \caption{Total number of switch points for up to 16 input ports in different planar designs. For non-planar designs~\cite{drost2016switching}, non-switching crosspoints are also taken into account. The x-axis represents size of input/output, y-axis represents the number of switches points and non switching crosspoints within the configurations.} 
    \label{fig:eval_disign_sw_count}
\end{figure}

In an ideal pairing mechanism, an arbitrary pair of input photons should traverse the same number of switch points in the same states.
This is generally not true in real configurations, leading to an imbalance between the number of switch points that each photon of a pair passes through.
In order to quantify this imbalance, we introduce the \textit{depth difference} $\Delta$, defined intuitively as the difference between the maximum and the minimum depths of the switch.
It is straightforward to see from Fig.~\ref{fig:triangular} that the minimum depth for the Triangular design is 0.
Therefore the $\Delta_{\text{triangular}} = N-2$, which is also the maximum depth.

For the Chevron design, the maximum depth depends on the parity of $N/2$.
When $N/2$ is odd, a photon traverses at most $N-3$ switch points, while its pairing partner needs to pass through at least $N/2-3$ switch points, giving $\Delta_{\text{chevron}}=N/2$.
The same depth difference is obtained for the case when $N/2$ is even. In this case, the maximum and minimum depths are given by $N-2$ and $N/2-2$, respectively.
In the Brickwork design, the maximum depth is $N/2$, while the minimum depth is $\lceil N/4 - 1 \rceil$, resulting $\Delta_{\text{brickwork}}=\lfloor N/4 + 1 \rfloor$.
\begin{table*}[htbp]
    \centering
    \begin{tabular}{| l || c | c | c | c |}
         %\cline{2-5}
         \hline
         & our designs  & Spanke \& Bene{\v{s}}~\cite{spanke1987n} & Bene{\v{s}}~\cite{benevs1964optimal} & Waksman ~\cite{waksman1968permutation} \\ \hline \hline
         Planar/non-planar & planar & planar & non-planar & non-planar \\ \hline
          No. of Switches & $N(N-2)/4$ & $N(N-1)/2$ & $N \log_2 -N/2   $ &  $N\log_2 N - N + 1$ \\ \hline
          Non-switching crosspoints  & $0$ & $0$ & $N(N-\log_2 N  - 1)/2 $ & $N(N-\log_2 N - 1 )/2 $ \\ \hline
         No. of coupling \& decoupling stages & $2$ & $2$ & $4\log_2 N - 2 $ & $4\log_2 N - 2 $ \\ \hline
    \end{tabular}
    \caption{ The resource complexity of different scalable designs for ($N \times N$) switching networks in terms of total number of switches, total number of cross points and total number of required interfaces for coupling and decoupling of the inputs.}
    \label{tab:Resource_Complexity}
\end{table*}

Out of the three proposed switches, the Brickwork design is the most balanced thanks to its lowest depth difference.
This means that the photons traversing the switch experience comparable losses.
The opposite is true for the Triangular design.
Whether this is an undesirable effect ultimately depends on the network traffic.
For quantum networks with fairly uniform traffic patterns, the Triangular design results in an uneven distribution of end-to-end Bell pair generation rates, leading to a decrease in the quality of service for certain connections.
On the other hand, we can envision scenarios where this imbalance may be a welcome feature.
The input ports with few switch points and their respective BSAs can be reserved for high-demand connections.
For example, the optical switch may be a component of a quantum gateway connecting two independent networks.
The low-loss input ports can be used for internetwork connections, while the more lossy input ports can be used for entanglement distribution within a single network.

\subsection{Comparison with other designs}
In this section, we compare the designs in terms of the number of switches with other related designs.
Fig.~\ref{fig:depth} illustrates the maximum depth for the our proposed designs and those analyzed in~\cite{spanke1987n}.
Given that our switches are designed for paired-egress BSA pools, it is not surprising that the their maximum depth is always lower than the optimal full permutational switch of~\cite{spanke1987n}.

Fig.~\ref{fig:eval_disign_sw_count} shows the minimum number of required switch points in related planar and non-planar designs for up to 16 input ports.
The focus of this paper has been planar designs suitable for chip fabrication, but of course non-planar designs can simply be ``flattened'' into planar form.
In this case, we have to consider the fixed crosspoints, which can be viewed as planar switch points with permanent \scross{} states.

Drost \emph{et al.}'s work proposing designs for small-scale non-planar switching networks is the work most comparable to our own, but does not address planar designs and does not include a scalable design~\cite{drost2016switching}.  Fig.~\ref{fig:eval_disign_sw_count} includes data on planarized Drost designs, counting switches and non-switching crosspoints up to 10 ports, the largest design they found.  More detailed follow-on design work needs to consider the non-planar design and analyze the photon loss based on the number of interfaces required for coupling and decoupling of input photons toward $2 \times 2$ switches.

Table~\ref{tab:Resource_Complexity} compares our proposed planar designs for paired egress to other scalable designs. To the best of our knowledge, no scalable structure for input pairing has been found before this work. Therefore we compare to some important full permutation switch networks.  As the table shows, the key result of our paper is the $>50\%$ reduction in switch points compared to Spanke and Bene{\v{s}}'s optimal planar design for full permutation switching networks~\cite{spanke1987n}.

\section{Conclusion}

We have shown that for an $N\times N$ switching network with paired-egress BSA Pools, the lower bound on the number of switching points in a planar architecture is $N(N-2)/4$.
We proposed three rearrangeable non-blocking designs that saturate this lower bound, along with their corresponding efficient routing algorithms.
Due to their recursive construction, our designs can be scaled to arbitrary size.

Our switch designs can be reversed in a straightforward manner.
The BSAs can be replaced with EPPS nodes, and the routing algorithms then distribute the entangled photon pairs to the desired outputs.
Therefore, our solution is directly applicable to the shared EPPS Pools switching problem considered in~\cite{drost2016switching}.

The shared BSA Pool architecture was recently used as an integral component in a proposal for distribution of remote entanglement between neutral ytterbium atoms coupled to optical cavities~\cite{li2024high}.
This demonstrates the relevance of optical switches with paired-egress BSA Pools, and the role they are expected to play in distributed quantum computing and quantum networking.

The performance of distributed systems crucially depends on how effectively we can harness the power of connected computers. This effectiveness largely relies on the performance of their network systems and architecture. Ideally, the network should always be capable of processing communication requests from computers; any delay waiting for network responses leads to decreased overall performance. Our rearrangeable non-blocking design ensures that the inputs and outputs of switches operate without stalling, achieving multiplexed parallel communication processing among different input-output pairs. Therefore, our optical switch designs prevent contention for communication from being a system-wide bottleneck. Thereby our work is essential for large-scale distributed quantum computers and the quantum Internet.

\appendices
\section*{Acknowledgment}
 The authors would like to thank Takao Tomono, Rikizo Ikuta, and Alto Osada for their help and fruitful discussions.

\bibliographystyle{IEEEtran-new.bst}

\bibliography{IEEEabrv, bibfile}

% Generated by IEEEtran.bst, version: 1.12 (2007/01/11)
\begin{thebibliography}{10}
\providecommand{\url}[1]{#1}
\csname url@samestyle\endcsname
\providecommand{\newblock}{\relax}
\providecommand{\bibinfo}[2]{#2}
\providecommand{\BIBentrySTDinterwordspacing}{\spaceskip=0pt\relax}
\providecommand{\BIBentryALTinterwordstretchfactor}{4}
\providecommand{\BIBentryALTinterwordspacing}{\spaceskip=\fontdimen2\font plus
\BIBentryALTinterwordstretchfactor\fontdimen3\font minus
  \fontdimen4\font\relax}
\providecommand{\BIBforeignlanguage}[2]{{%
\expandafter\ifx\csname l@#1\endcsname\relax
\typeout{** WARNING: IEEEtran.bst: No hyphenation pattern has been}%
\typeout{** loaded for the language `#1'. Using the pattern for}%
\typeout{** the default language instead.}%
\else
\language=\csname l@#1\endcsname
\fi
#2}}
\providecommand{\BIBdecl}{\relax}
\BIBdecl

\bibitem{clos1953study}
C.~Clos, ``A study of non-blocking switching networks,'' \emph{Bell System
  Technical Journal}, vol.~32, no.~2, pp. 406--424, 1953,
  \href{https://dx.doi.org/10.1002/j.1538-7305.1953.tb01433.x}{doi:10.1002/j.1538-7305.1953.tb01433.x}.

\bibitem{opferman1971class}
D.~C. Opferman and N.~T. Tsao-Wu, ``On a class of rearrangeable switching
  networks part i: Control algorithm,'' \emph{The Bell System Technical
  Journal}, vol.~50, no.~5, pp. 1579--1600, 1971,
  \href{https://dx.doi.org/10.1002/j.1538-7305.1971.tb02569.x}{doi:10.1002/j.1538-7305.1971.tb02569.x}.

\bibitem{benevs1965mathematical}
V.~E. Bene{\v{s}}, \emph{Mathematical theory of connecting networks and
  telephone traffic}.\hskip 1em plus 0.5em minus 0.4em\relax Academic press,
  1965, \href{https://dx.doi.org/10.2307/2343633}{doi:10.2307/2343633}.

\bibitem{benevs1964optimal}
V.~E. Beneš, ``Optimal rearrangeable multistage connecting networks,''
  \emph{The Bell System Technical Journal}, vol.~43, no.~4, pp. 1641--1656,
  1964,
  \href{https://dx.doi.org/10.1002/j.1538-7305.1964.tb04103.x}{doi:10.1002/j.1538-7305.1964.tb04103.x}.

\bibitem{waksman1968permutation}
A.~Waksman, ``A permutation network,'' \emph{Journal of the ACM (JACM)},
  vol.~15, no.~1, pp. 159--163, 1968,
  \href{https://dx.doi.org/10.1145/321439.321449}{doi:10.1145/321439.321449}.

\bibitem{athas:multicomputer}
W.~C. Athas and C.~L. Seitz, ``Multicomputers: message-passing concurrent
  computers,'' \emph{IEEE Computer}, vol.~21, pp. 9--24, Aug. 1988,
  \href{https://dx.doi.org/10.1109/2.73}{doi:10.1109/2.73}.

\bibitem{dally04:_interconnects}
W.~J. Dally and B.~Towles, \emph{Principles and Practices of Interconnection
  Networks}.\hskip 1em plus 0.5em minus 0.4em\relax Elsevier, 2004.

\bibitem{awschalom2022roadmap}
D.~D. Awschalom \emph{et~al.}, ``A roadmap for quantum interconnects,'' Argonne
  National Laboratory (ANL), Argonne, IL (United States), Tech. Rep., 2022,
  \href{https://dx.doi.org/10.2172/1900586}{doi:10.2172/1900586}.

\bibitem{jiang07:PhysRevA.76.062323}
\BIBentryALTinterwordspacing
L.~Jiang, J.~M. Taylor, A.~S. S\o{}rensen, and M.~D. Lukin, ``Distributed
  quantum computation based on small quantum registers,'' \emph{Phys. Rev. A},
  vol.~76, p. 062323, Dec 2007,
  \href{https://dx.doi.org/10.1103/PhysRevA.76.062323}{doi:10.1103/PhysRevA.76.062323}.
\BIBentrySTDinterwordspacing

\bibitem{jiang2010scalable}
L.~Jiang, J.~M. Taylor, A.~S. S{\o}rensen, and M.~D. Lukin, ``Scalable quantum
  networks based on few-qubit registers,'' \emph{International Journal of
  Quantum Information}, vol.~8, no. 01n02, pp. 93--104, 2010,
  \href{https://dx.doi.org/10.1142/S0219749910006058}{doi:10.1142/S0219749910006058}.

\bibitem{kim09:_integ_optic_ion_trap}
J.~Kim and C.~Kim, ``Integrated optical approach to trapped ion quantum
  computation,'' \emph{QIC}, vol.~9, no.~2, 2009.

\bibitem{kim05:_system}
J.~Kim \emph{et~al.}, ``System design for large-scale ion trap quantum
  information processor,'' \emph{QIC}, vol.~5, no.~7, pp. 515--537, 2005,
  \href{https://dx.doi.org/10.26421/QIC5.7-1}{doi:10.26421/QIC5.7-1}.

\bibitem{lim05:_repeat_until_success}
Y.~L. Lim, S.~D. Barrett, A.~Beige, P.~Kok, and L.~C. Kwek,
  ``{Repeat-Until-Success} quantum computing using stationary and flying
  qubits,'' \emph{Physical Review Letters}, vol.~95, no.~3, p. 30505, 2005,
  \href{https://dx.doi.org/10.1103/PhysRevA.73.012304}{doi:10.1103/PhysRevA.73.012304}.

\bibitem{PhysRevA.89.022317}
\BIBentryALTinterwordspacing
C.~Monroe \emph{et~al.}, ``Large-scale modular quantum-computer architecture
  with atomic memory and photonic interconnects,'' \emph{Phys. Rev. A},
  vol.~89, p. 022317, Feb 2014,
  \href{https://dx.doi.org/10.1103/PhysRevA.89.022317}{doi:10.1103/PhysRevA.89.022317}.
\BIBentrySTDinterwordspacing

\bibitem{nickerson2013topological}
N.~H. Nickerson, Y.~Li, and S.~C. Benjamin, ``Topological quantum computing
  with a very noisy network and local error rates approaching one percent,''
  \emph{Nature communications}, vol.~4, p. 1756, 2013,
  \href{https://dx.doi.org/10.1038/ncomms2773}{doi:10.1038/ncomms2773}.

\bibitem{oi06:_dist-ion-trap-qec}
D.~K.~L. Oi, S.~J. Devitt, and L.~C.~L. Hollenberg, ``Scalable error correction
  in distributed ion trap computers,'' \emph{Physical Review A}, vol.~74, p.
  052313, 2006,
  \href{https://dx.doi.org/10.48550/arXiv.quant-ph/0606226}{doi:10.48550/arXiv.quant-ph/0606226}.

\bibitem{van-meter06:thesis}
R.~Van{ }Meter{ }III, ``Architecture of a quantum multicomputer optimized for
  {Shor's} factoring algorithm,'' Ph.D. dissertation, Keio University, 2006,
  available as arXiv:quant-ph/0607065.
  \href{https://dx.doi.org/10.48550/arXiv.quant-ph/0607065}{doi:10.48550/arXiv.quant-ph/0607065}.

\bibitem{van-meter16:_ieee-comp}
R.~Van{ }Meter and S.~Devitt, ``The path to scalable distributed quantum
  computing,'' \emph{IEEE Computer}, vol.~49, no.~9, pp. 31--42, Sep. 2016,
  \href{https://dx.doi.org/10.1109/MC.2016.291}{doi:10.1109/MC.2016.291}.

\bibitem{yimsiriwattana2004distributed}
A.~Yimsiriwattana and S.~J. Lomonaco~Jr, ``Distributed quantum computing: A
  distributed shor algorithm,'' in \emph{Quantum Information and Computation
  II}, vol. 5436.\hskip 1em plus 0.5em minus 0.4em\relax SPIE, 2004, pp.
  360--372, \href{https://dx.doi.org/10.1117/12.546504}{doi:10.1117/12.546504}.

\bibitem{diadamo21:dist-vqe}
\BIBentryALTinterwordspacing
S.~DiAdamo, M.~Ghibaudi, and J.~Cruise, ``Distributed quantum computing and
  network control for accelerated {VQE},'' \emph{IEEE Transactions on Quantum
  Engineering}, vol.~2, pp. 1--21, 2021,
  \href{https://dx.doi.org/10.1109/TQE.2021.3057908}{doi:10.1109/TQE.2021.3057908}.
\BIBentrySTDinterwordspacing

\bibitem{satoh2020federated}
\BIBentryALTinterwordspacing
R.~Satoh, M.~Hajdušek, and R.~Van~Meter, ``Federated graph state preparation
  on noisy, distributed quantum computers,'' \emph{IPSJ Quantum Software SIG
  Technical Reports}, vol. 2020-QS-1, p.~6, 2020.
\BIBentrySTDinterwordspacing

\bibitem{van2006distributed}
R.~Van~Meter, K.~Nemoto, W.~Munro, and K.~M. Itoh, ``Distributed arithmetic on
  a quantum multicomputer,'' \emph{ACM SIGARCH Computer Architecture News},
  vol.~34, no.~2, pp. 354--365, 2006,
  \href{https://dx.doi.org/10.48550/arXiv.quant-ph/0607160}{doi:10.48550/arXiv.quant-ph/0607160}.

\bibitem{van2007communication}
R.~Van~Meter, K.~Nemoto, and W.~Munro, ``Communication links for distributed
  quantum computation,'' \emph{IEEE Transactions on Computers}, vol.~56,
  no.~12, pp. 1643--1653, 2007,
  \href{https://dx.doi.org/10.48550/arXiv.quant-ph/0701043}{doi:10.48550/arXiv.quant-ph/0701043}.

\bibitem{caleffi2022distributed}
M.~Caleffi \emph{et~al.}, ``Distributed quantum computing: a survey,''
  \emph{arXiv preprint arXiv:2212.10609}, 2022,
  \href{https://dx.doi.org/10.48550/arXiv.2212.10609}{doi:10.48550/arXiv.2212.10609}.

\bibitem{gauthier2023architecture}
S.~Gauthier, G.~Vardoyan, and S.~Wehner, ``An architecture for control of
  entanglement generation switches in quantum networks,'' \emph{IEEE
  Transactions on Quantum Engineering}, vol.~4, no.~01, pp. 1--17, 2023,
  \href{https://dx.doi.org/10.1109/TQE.2023.3320047}{doi:10.1109/TQE.2023.3320047}.

\bibitem{alshowkan2012reconfigurable}
\BIBentryALTinterwordspacing
M.~Alshowkan \emph{et~al.}, ``Reconfigurable quantum local area network over
  deployed fiber,'' \emph{PRX Quantum}, vol.~2, p. 040304, Oct 2021,
  \href{https://dx.doi.org/10.1103/PRXQuantum.2.040304}{doi:10.1103/PRXQuantum.2.040304}.
\BIBentrySTDinterwordspacing

\bibitem{bersin2024development}
\BIBentryALTinterwordspacing
E.~Bersin \emph{et~al.}, ``Development of a {B}oston-area 50-km fiber quantum
  network testbed,'' \emph{Phys. Rev. Appl.}, vol.~21, p. 014024, Jan 2024,
  \href{https://dx.doi.org/10.1103/PhysRevApplied.21.014024}{doi:10.1103/PhysRevApplied.21.014024}.
\BIBentrySTDinterwordspacing

\bibitem{kozlowski2023rfc}
W.~Kozlowski \emph{et~al.}, ``Rfc 9340: Architectural principles for a quantum
  internet,'' 2023,
  \href{https://dx.doi.org/10.17487/RFC9340}{doi:10.17487/RFC9340}.

\bibitem{hajduvsek2023quantum}
M.~Hajdu{\v{s}}ek and R.~Van~Meter, ``Quantum communications,'' \emph{arXiv
  preprint arXiv:2311.02367}, 2023,
  \href{https://dx.doi.org/10.48550/arXiv.2311.02367}{doi:10.48550/arXiv.2311.02367}.

\bibitem{wehner2018quantum}
\BIBentryALTinterwordspacing
S.~Wehner, D.~Elkouss, and R.~Hanson, ``Quantum internet: A vision for the road
  ahead,'' \emph{Science}, vol. 362, no. 6412, p. eaam9288, 2018,
  \href{https://dx.doi.org/10.1126/science.aam9288}{doi:10.1126/science.aam9288}.
\BIBentrySTDinterwordspacing

\bibitem{vanmeter2022quantum}
R.~Van~Meter \emph{et~al.}, ``A quantum internet architecture,'' in \emph{2022
  IEEE International Conference on Quantum Computing and Engineering (QCE)},
  2022, pp. 341--352,
  \href{https://dx.doi.org/10.1109/QCE53715.2022.00055}{doi:10.1109/QCE53715.2022.00055}.

\bibitem{van2014quantum}
R.~Van~Meter, \emph{Quantum Networking}.\hskip 1em plus 0.5em minus 0.4em\relax
  John Wiley \& Sons, 2014,
  \href{https://dx.doi.org/10.1002/9781118648919}{doi:10.1002/9781118648919}.

\bibitem{drost2016switching}
R.~J. Drost, T.~J. Moore, and M.~Brodsky, ``Switching networks for
  pairwise-entanglement distribution,'' \emph{Journal of Optical Communications
  and Networking}, vol.~8, no.~5, pp. 331--342, 2016,
  \href{https://dx.doi.org/10.1364/JOCN.8.000331}{doi:10.1364/JOCN.8.000331}.

\bibitem{spanke1987n}
\BIBentryALTinterwordspacing
R.~A. Spanke and V.~Bene{\v{s}}, ``N-stage planar optical permutation
  network,'' \emph{Applied Optics}, vol.~26, no.~7, pp. 1226--1229, 1987,
  \href{https://dx.doi.org/10.1364/AO.26.001226}{doi:10.1364/AO.26.001226}.
\BIBentrySTDinterwordspacing

\bibitem{krutyanskyi2023entanglement}
\BIBentryALTinterwordspacing
V.~Krutyanskiy \emph{et~al.}, ``Entanglement of trapped-ion qubits separated by
  230 meters,'' \emph{Phys. Rev. Lett.}, vol. 130, p. 050803, Feb 2023,
  \href{https://dx.doi.org/10.1103/PhysRevLett.130.050803}{doi:10.1103/PhysRevLett.130.050803}.
\BIBentrySTDinterwordspacing

\bibitem{crespi2011integrated}
A.~Crespi \emph{et~al.}, ``Integrated photonic quantum gates for polarization
  qubits,'' \emph{Nature communications}, vol.~2, no.~1, p. 566, 2011,
  \href{https://dx.doi.org/10.1038/ncomms1570}{doi:10.1038/ncomms1570}.

\bibitem{corrielli2014rotated}
G.~Corrielli \emph{et~al.}, ``Rotated waveplates in integrated waveguide
  optics,'' \emph{Nature communications}, vol.~5, no.~1, p. 4249, 2014,
  \href{https://dx.doi.org/10.1038/ncomms5249}{doi:10.1038/ncomms5249}.

\bibitem{xu2022self}
X.~Xu \emph{et~al.}, ``Self-calibrating programmable photonic integrated
  circuits,'' \emph{Nature Photonics}, vol.~16, no.~8, pp. 595--602, 2022,
  \href{https://dx.doi.org/10.1038/s41566-022-01020-z}{doi:10.1038/s41566-022-01020-z}.

\bibitem{bogaerts2020programmable}
W.~Bogaerts \emph{et~al.}, ``Programmable photonic circuits,'' \emph{Nature},
  vol. 586, no. 7828, pp. 207--216, 2020,
  \href{https://dx.doi.org/10.1038/s41586-020-2764-0}{doi:10.1038/s41586-020-2764-0}.

\bibitem{wang2020integrated}
J.~Wang, F.~Sciarrino, A.~Laing, and M.~G. Thompson, ``Integrated photonic
  quantum technologies,'' \emph{Nature Photonics}, vol.~14, no.~5, pp.
  273--284, 2020,
  \href{https://dx.doi.org/10.1038/s41566-019-0532-1}{doi:10.1038/s41566-019-0532-1}.

\bibitem{saleh2019fundamentals}
B.~E. Saleh and M.~C. Teich, \emph{Fundamentals of Photonics}.\hskip 1em plus
  0.5em minus 0.4em\relax John Wiley \& Sons, 2019,
  \href{https://dx.doi.org/10.1002/0471213748}{doi:10.1002/0471213748}.

\bibitem{kim03:_1100_port_mems}
J.~Kim \emph{et~al.}, ``1100x1100 port {MEMS}-based optical crossconnect with
  {4-dB} maximum loss,'' \emph{IEEE Photonics Technology Letters}, vol.~15,
  no.~11, pp. 1537--1539, 2003,
  \href{https://dx.doi.org/10.1109/LPT.2003.818653}{doi:10.1109/LPT.2003.818653}.

\bibitem{harris2016large}
N.~C. Harris \emph{et~al.}, ``Large-scale quantum photonic circuits in
  silicon,'' \emph{Nanophotonics}, vol.~5, no.~3, pp. 456--468, 2016,
  \href{https://dx.doi.org/10.1515/nanoph-2015-0146}{doi:10.1515/nanoph-2015-0146}.

\bibitem{lonvcar2000waveguiding}
M.~Lon{\v{c}}ar, D.~Nedeljkovi{\'c}, T.~Doll, J.~Vu{\v{c}}kovi{\'c},
  A.~Scherer, and T.~P. Pearsall, ``Waveguiding in planar photonic crystals,''
  \emph{Applied Physics Letters}, vol.~77, no.~13, pp. 1937--1939, 2000,
  \href{https://dx.doi.org/10.1063/1.1311604}{doi:10.1063/1.1311604}.

\bibitem{o2009photonic}
J.~L. O'Brien, A.~Furusawa, and J.~Vu{\v{c}}kovi{\'c}, ``Photonic quantum
  technologies,'' \emph{Nature photonics}, vol.~3, no.~12, pp. 687--695, 2009,
  \href{https://dx.doi.org/10.48550/arXiv.1003.3928}{doi:10.48550/arXiv.1003.3928}.

\bibitem{taballione20198}
C.~Taballione \emph{et~al.}, ``8$\times$ 8 reconfigurable quantum photonic
  processor based on silicon nitride waveguides,'' \emph{Optics express},
  vol.~27, no.~19, pp. 26\,842--26\,857, 2019,
  \href{https://dx.doi.org/10.1364/OE.27.026842}{doi:10.1364/OE.27.026842}.

\bibitem{nesic2019photonic}
A.~Nesic \emph{et~al.}, ``Photonic-integrated circuits with non-planar
  topologies realized by 3d-printed waveguide overpasses,'' \emph{Optics
  Express}, vol.~27, no.~12, pp. 17\,402--17\,425, 2019,
  \href{https://dx.doi.org/10.1364/OE.27.017402}{doi:10.1364/OE.27.017402}.

\bibitem{li2024high}
\BIBentryALTinterwordspacing
Y.~Li and J.~Thompson, ``High-rate and high-fidelity modular interconnects
  between neutral atom quantum processors,'' \emph{arXiv preprint
  arXiv:2401.04075}, 2024,
  \href{https://dx.doi.org/10.48550/arXiv.2401.04075}{doi:10.48550/arXiv.2401.04075}.
\BIBentrySTDinterwordspacing

\end{thebibliography}

\end{document}